\documentstyle[twoside,epsf]{article}

\input ibvs2.sty

\begin{document}

\IBVShead{5477}{14 November 2003}

\IBVStitletl{Optical CCD Observations of Eta Carinae}{ at La Plata Observatory}

\IBVSauth{Fern\'andez Laj\'us, E.$^{1,2}$; Gamen, R.$^{1,3}$; Schwartz, M.$^1$; Salerno, N.$^1$; Llinares, C.$^1$; Fari\~na, C.$^1$; Amor\'{\i}n, R.$^1$; Niemela, V.$^{1,4}$}

\IBVSinst{Facultad de Ciencias Astron\'omicas y Geof\'{\i}sicas, Universidad Nacional de La Plata, Paseo del Bosque S/N. B1900FWA La Plata.}
\IBVSinst{Fellow of UNLP, La Plata, Argentina; e-mail: eflajus@fcaglp.unlp.edu.ar}
\IBVSinst{Fellow of CONICET, Argentina.}
\IBVSinst{Member of the Carrera del Investigador Cient\'{\i}fico, CIC, Buenos Aires, Argentina.}

\SIMBADobjAlias{Eta Car}{HD 93308}
\IBVStyp{ LBV }

\IBVSabs{In 2003.5 Eta Carinae was expected to undergo an X-ray eclipse}
\IBVSabs{(Damineli et al., 2000). In the framework of an international}
\IBVSabs{campaign to obtain multi-wavelength observations of this event,}
\IBVSabs{we have obtained optical CCD images of Eta Carinae. Here, we present}
\IBVSabs{the B, V, R, I, data of Eta Car obtained before}
\IBVSabs{and during the X-ray eclipse.}

\begintext

In 2003.5 Eta Carinae, suspected to be a binary system with a period of 
5.53 years (cf. Damineli et al., 2000),
was expected to undergo an X-ray eclipse (cf. Corcoran et al., 2001). 
In the framework of an international campaign to obtain multi-wavelength 
observations of this event, we have obtained optical CCD images of Eta Carinae. 
About 3000 images were acquired in 2003 between January and August.
Here we present our data of Eta Car obtained before and during the 
X-ray eclipse. 

\IBVSfig{8cm}{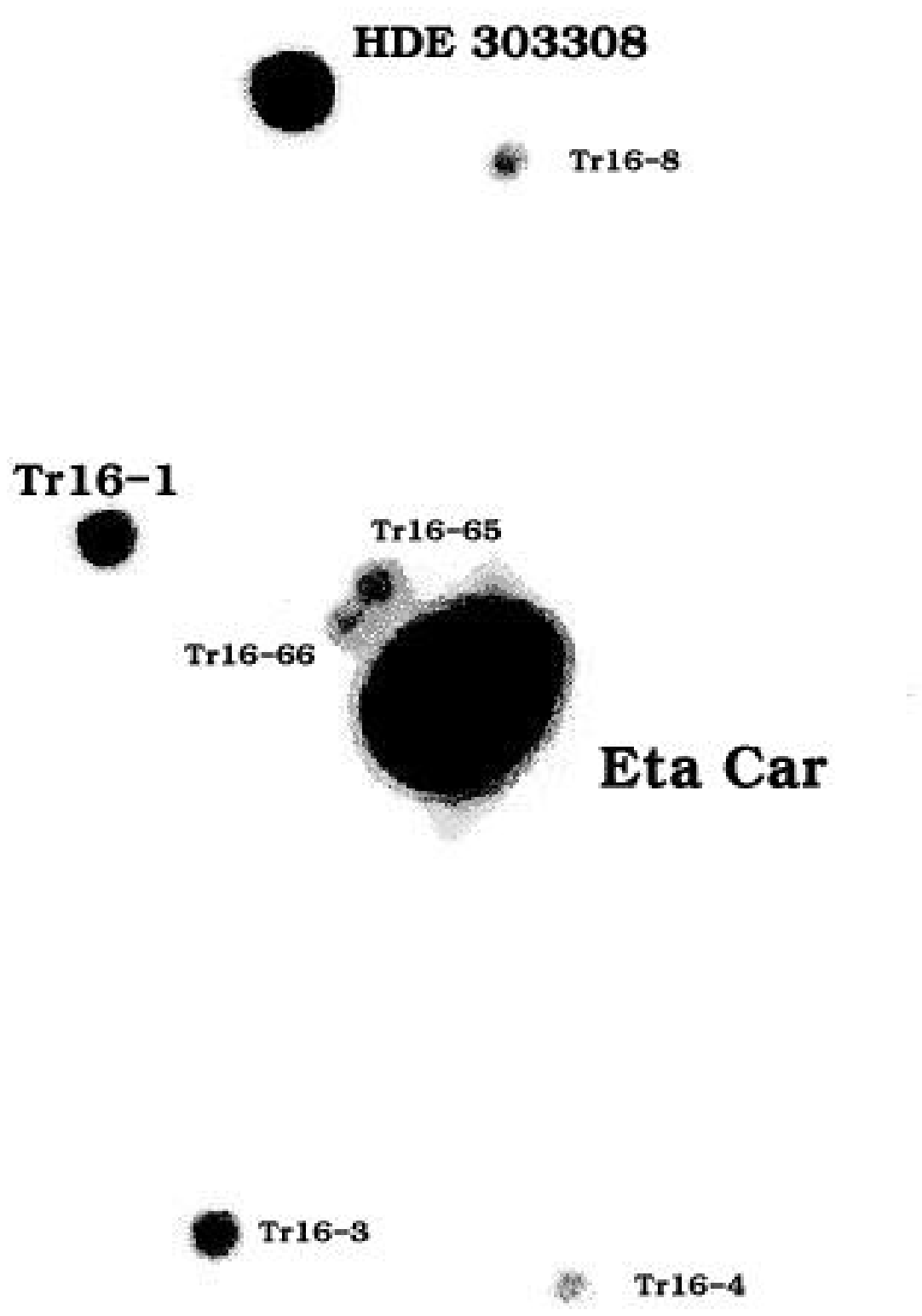}{$V$ image of the Eta Carinae region. 
North is up and East is to the left. Labels follow the nomenclature of 
Feinstein et al. (1973).}
%\IBVSfigKey{5477-f1}{Eta Car}{finding chart}

\clearpage

\IBVSfig{16cm}{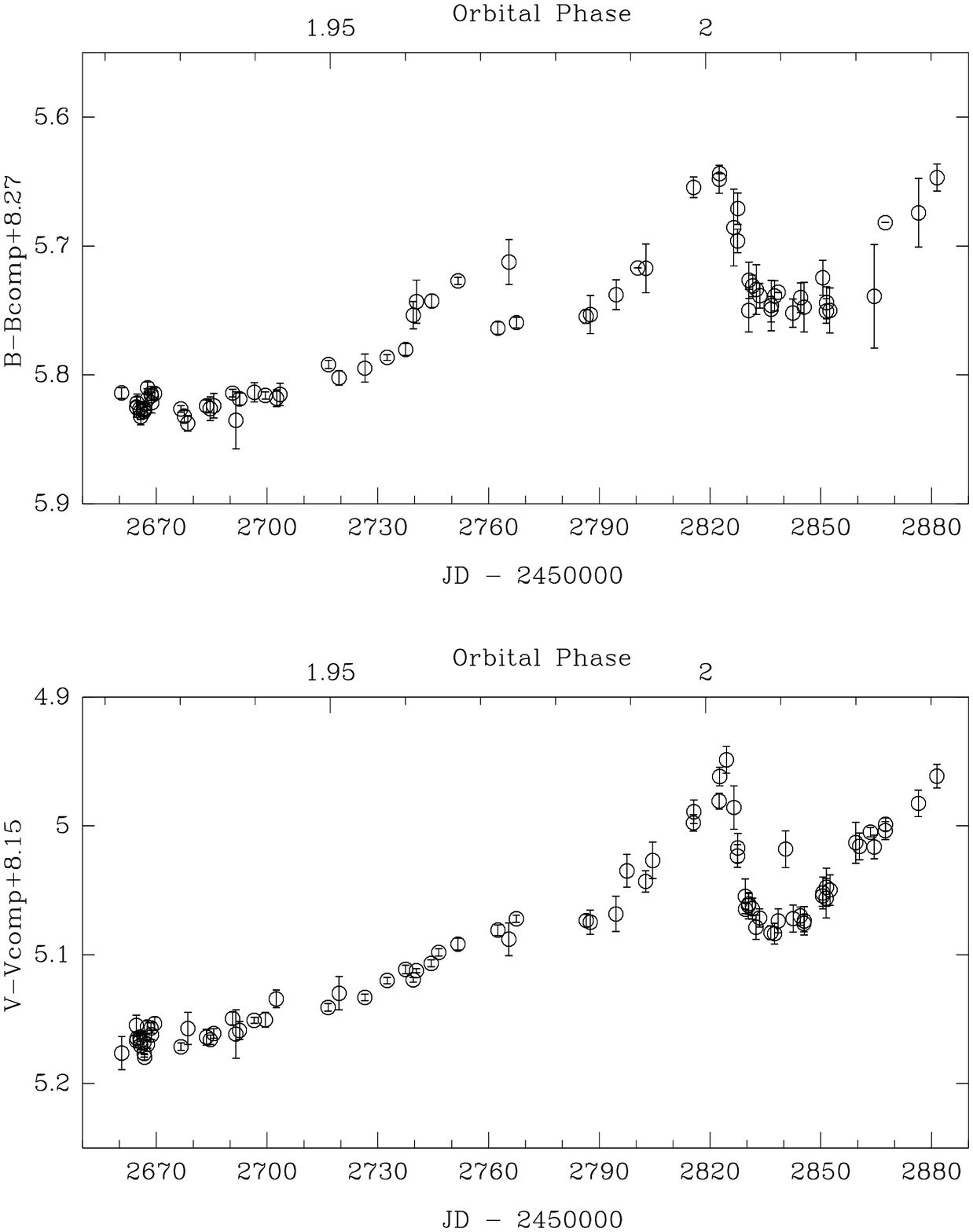}{Relative $B$ and $V$ 
magnitudes of Eta Car observed between 20th January and
29th August, 2003. Bars represent the sample standard deviations.
Along the top axis, orbital phases are indicated,
calculated according to the ephemeris:
heliocentric X-ray minimum = 1997.95 + 5.53609E (Corcoran 2003).}
%\IBVSfigKey{5477-f2.eps}{Eta Car}{light curve}

The images were acquired through Johnson-Cousins {\em BVRI} filters 
with a Photometrics STAR I CCD camera attached to the 0.8-m reflector 
(f/20.06 Cassegrain) at La Plata Observatory, Argentina. 
The detector used is a 
Thomson-CSF TH7883 scientific-grade front-illuminated chip, 
Peltier cooled, of 384 $\times$ 576 pixels (23 $\mu$m square pixel).
Our instrumental configuration results in 1\farcm9 $\times$ 2\farcm8 
field images with an oversampled scale of 0\farcs3 per pixel.
{\em BVRI} passbands used are those recommended by Bessell (1990)
for coated CCDs. One of our images is reproduced in 
Figure~\ref{fig:5477-f1.eps}, with identifications of the objects.

Differential photometry of Eta Car was determined using HDE~303308 as
comparison star. This star was found to have constant light by Sterken 
et al (2001) and Freyhammer et al (2001). In order to give values approximate 
to the standard magnitudes of Eta Car, we have added to our relative 
magnitudes the $UBVRI$ Johnson-Kron-Cousins photometry of HDE~303308 
by Feinstein (1982), i.e. $B$ = 8.27, $V$ = 8.15, $R$ = 8.01 and $I$ = 7.85.

Instrumental magnitudes of each star were determined by means of aperture 
photometry. In order to minimize small fluctuations, due to noise, 
the instrumental magnitudes, were calculated as an average
of individual values determined in 6 apertures for Eta Car and 4 apertures
for  HDE~303308, selected constructing CCD growth
curves (Howell et al., 1989) for the first observed frames. The radii of
the apertures for Eta Car were between 80 and 105 pixels, and those for
 HDE~303308 between 40 and 60 pixels, with increments of 5 pixels. The values
of the apertures were kept the same for all of the frames observed during the
campaign.

\IBVSfig{16cm}{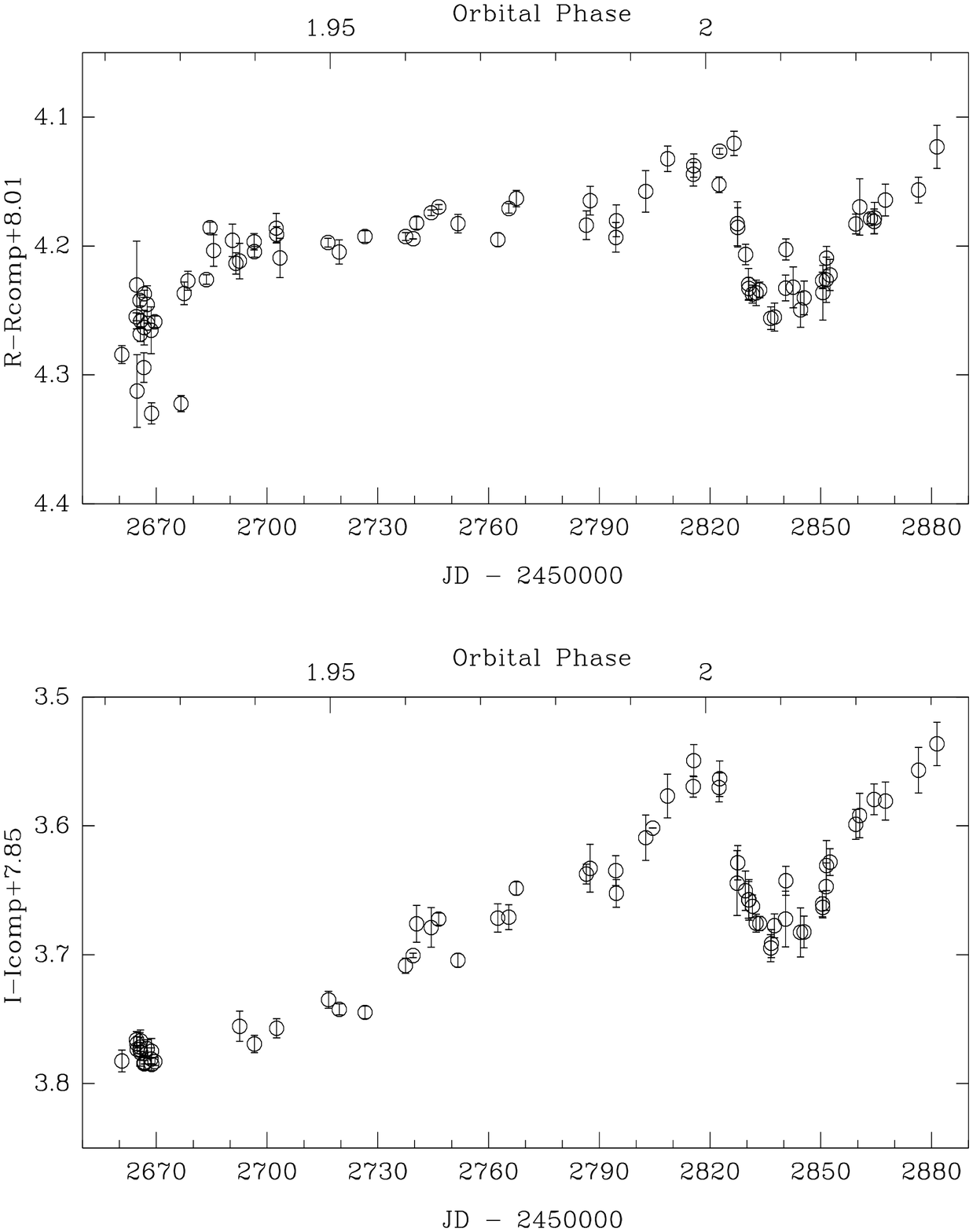}{The same as Fig.~2 for relative $R$ and $I$ 
magnitudes.}
%\IBVSfigKey{5477-f3.eps}{Eta Car}{light curve}

Typical individual errors of {\em B, V, R, I} instrumental
magnitudes are 0.008, 0.004, 0.006, and 0.01 mag respectively.
Mean differential magnitudes of Eta Car were calculated from the instrumental
magnitudes of a series 
of images obtained within intervals of 100-minutes ($\sim$ 0.000035 
of orbital phase). Standard deviations were also calculated for each 
mean magnitude.
These data are available through the IBVS-website as a table 
({\tt 5477-t1.txt}) where, in successive columns we quote for each 
filter, the Julian Date of each observation, the mean differential 
magnitude, the standard deviation of the mean, and the number of 
images included in the mean values.

\IBVSedata{5477-t1.txt}
%\IBVSdataKey{5477-t1.txt}{Eta Car}{photometry}

Variations of the observed {\em B, V, R, I} magnitudes of Eta Car 
from January to August, 2003, are shown in Figures~\ref{fig:5477-f2.eps} 
and~\ref{fig:5477-f3.eps}.
The optical variations seem to follow the behaviour of Eta Car observed in 
X-rays by RXTE, available on the web page of Dr. M. Corcoran (2003).
We notice a fading of optical light about 10 days after the eclipse in X-ray
was observed (phase 2.0 in the Figures~\ref{fig:5477-f2.eps}  
and~\ref{fig:5477-f3.eps}). This fading is similar in shape to the
infrared minimum observed during the previous X-ray eclipse of Eta Car
(Feast et al., 2001).

An interpretation of this light fading as an optical eclipse of the
Eta Car system is pending an analysis incorporating all the data collected 
during the multi-wavelength campaign of observations of this 2003.5 event.

The participation of the following students in obtaining the observations: 
Anabella Araudo, Gisela Romero, Ariel S\'anchez Camus,
Silvia Sicilia, Lautaro Simontacchi, Andrea Torres and Javier V\'asquez
is gratefully acknowledged.
We thank the referee, Dr. C. Sterken, for valuable suggestions which improved
the presentation of this paper.

\references

Bessell, M., 1990, {\it PASP}, {\bf 102}, 1181 
%%\BIBCODE{1990PASP..102.1181B} 

Corcoran, M. F., Ishibashi, K., Swank, J. H., \& Petre, R., 2001, {\it ApJ}, {\bf 547}, 1034 
%%\BIBCODE{2001ApJ...547.1034C}

Corcoran, M. F., 2003, private communication, see \hfill\break
http://lheawww.gsfc.nasa.gov/users/corcoran/eta\_car/etacar\_rxte\_lightcurve/

Damineli, A., Kaufer, A., Wolf, B., Stahl, O., Lopes, D., \& de Ara\'ujo, F.,
2000, {\it ApJ}, {\bf 528}, L101 
%%\BIBCODE{2000ApJ...528L.101D}

Feast, M., Whitelock, P., Marang, F., 2001, {\it MNRAS}, {\bf 322}, 741
%%\BIBCODE{2001MNRAS.322..741F}
 
Feinstein, A., 1982, {\it AJ}, {\bf 87}, 1012 
%%\BIBCODE{1982AJ.....87.1012F}

Feinstein, A., Marraco, H., Muzzio, J.C., 1973, {\it A\&AS}, {\bf 12}, 331
%%\BIBCODE{1973A&AS...12..331F}

Freyhammer, L., Clausen, J., Arentoft, T., \& Sterken, C., 2001, {\it A\&A}, {\bf 369}, 561 
%%\BIBCODE{2001A+A...369..561F}

Howell, S. B., 1989, {\it PASP}, {\bf 101}, 616 
%%\BIBCODE{1989PASP..101..616H}

Sterken, C., Freyhammer, L., Arentoft, T., van Genderen, A. M., 2001, {\it ASP Conf Ser}, {\bf 233}, 39 
%%\BIBCODE{2001pcyg.conf...39S}

\endreferences 

\end{document}